\documentclass[floatfix,%
aps,
twocolumn,
amsmath,amssymb,
]{revtex4-1}

\usepackage{graphicx}% Include figure files
\usepackage{dcolumn}% Align table columns on decimal point
\usepackage{bm}% bold math
\usepackage[T1]{fontenc}
\usepackage{mathptmx}
\usepackage{etoolbox}
\usepackage{multirow}
\usepackage{booktabs}
\usepackage{tabularx}
\usepackage{lipsum}
\usepackage{soul}
\makeatletter
\def\@email#1#2{%
 \endgroup
 \patchcmd{\titleblock@produce}
  {\frontmatter@RRAPformat}
  {\frontmatter@RRAPformat{\produce@RRAP{*#1\href{mailto:#2}{#2}}}\frontmatter@RRAPformat}
  {}{}
}%
\makeatother
\usepackage[version=3]{mhchem} % Formula subscripts using \ce{}
\usepackage{color}
\usepackage{graphicx}
\usepackage{xcolor}
\usepackage{comment}

\usepackage{hyperref}
\hypersetup{colorlinks=true}
\usepackage[all]{hypcap} % let hyperlinks correctly point to figures 

\DeclareUnicodeCharacter{0301}{\'{e}} 

\begin{document}

%\preprint{AIP/123-QED}
%%%%%%%%%%%%%%%%%%%%%%%%%%%%%%%%%%%%%%%%%%%%%%%%%%%%%%%%%%%%%%%%%%%%%
\title{Heat transport at the nanoscale and ultralow temperatures - implications for quantum technologies}%%%%%%%%%%%%%%%%%%%%%%%%%%%%%%%%%%%%%%%%%%%%%%%%%%%%%%%%%%%%%%%%%%%
\author{Danial Majidi}
\affiliation{\mbox{Univ.} Grenoble Alpes, CNRS, LPMMC, 25 rue des Martyrs, Grenoble, France}

\author{Justin P. Bergfield}
\affiliation{Department of Physics, Illinois State University, Normal, IL 61790, USA}

%\author{Johannes H\"ofer}
%\affiliation{\mbox{Univ.} Grenoble Alpes, CNRS, Grenoble INP, Institut N\'eel, 25 rue des Martyrs, Grenoble, France}

\author{Ville Maisi}
\affiliation{NanoLund and Solid State Physics, Lund University, Box 118, 22100 Lund, Sweden}

\author{Johannes H\"ofer}
\affiliation{\mbox{Univ.} Grenoble Alpes, CNRS, Grenoble INP, Institut N\'eel, 25 rue des Martyrs, Grenoble, France}

\author{Herv\'e Courtois}
\affiliation{\mbox{Univ.} Grenoble Alpes, CNRS, Grenoble INP, Institut N\'eel, 25 rue des Martyrs, Grenoble, France}

\author{Clemens B. Winkelmann}
\affiliation{\mbox{Univ.} Grenoble Alpes, CNRS, Grenoble INP, Institut N\'eel, 25 rue des Martyrs, Grenoble, France}
\affiliation{\mbox{Univ.} Grenoble Alpes, CEA, Grenoble INP, IRIG-Pheliqs, Grenoble, France.}

\email{clemens.winkelmann@grenoble-inp.fr}

\date{\today}% It is always \today, today,

\begin{abstract}
In this perspective, we discuss thermal imbalance and the associated electron-mediated thermal transport in quantum electronic devices at very low temperatures. We first present the theoretical approaches describing heat transport in nanoscale conductors at low temperatures, in which quantum confinement and interactions play an important role. We then discuss the experimental techniques for generating and measuring heat currents and temperature gradients on the nanoscale. Eventually we review the most important quantum effects on heat transport, and discuss implications for quantum technologies and future directions in the field.
\end{abstract}

%\abbreviations{IR,NMR,UV}
%\keywords{.....}
\maketitle
%\tableofcontents

%%%%%%%%%%%%%%%%%%%%%%%%%%%%%%%%%%%%%%%%%%%%%%%%%%%%%%%%%%%%%%%%%%%%%
%% The manuscript does not need to include \maketitle, which is
%% executed automatically. The document should begin with an
%% abstract, if appropriate. If one is given and should not be, the
%% contents will be gobbled.
%%%%%%%%%%%%%%%%%%%%%%%%%%%%%%%%%%%%%%%%%%%%%%%%%%%%%%%%%%%%%%%%%%%%%

\section{Introduction}

Heat transport and dissipation have emerged as important issues related to the performance of quantum devices. In particular, the unavoidable heat release in quantum bits under operation is extremely localized at a nanometric scale. Such localized heating is now understood to detrimentally affect the fidelity of quantum computation, with local power dissipation emerging as a primary constraint on quantum coherence, even in processors with a few physical qubits \cite{philips2022universal, undseth_hotter_2023}.

Realizing efficient heat transfer at the quantum scale is therefore critical, requiring a detailed understanding of heat transport in open quantum systems. At low temperatures, the thermal couplings can be extremely weak, leading to populations of electrons, phonons, photons, spins and other excitations coexisting at different temperatures within the same structure \cite{giazotto2006opportunities}. At low temperature, the faster decrease, compared to their electronic counterpart, of both the photon and phonon densities and the coupling to these modes makes the usual bosonic heat transport channels highly inefficient so that electronic heat transport dominates.

For a free-electron gas, the Wiedemann-Franz (WF) law states that the ratio between the electronic charge and thermal conductances $G$ and $\kappa$, respectively, is given by the temperature. It is expressed as $\kappa/G = L_0 T$, where $L_0=(\pi^2/3)(k_B/e)^2$ is the Lorentz number, $T$ is the temperature, $e$ the elementary charge and $k_B$ is Boltzmann's constant. Empirically, the WF law is widely obeyed to a semi-quantitative level in most electrically conducting materials, such that deviations can be an excellent marker of unconventional or non-classical transport mechanisms \cite{crossno2016,lee2017anomalously,bergfield2009thermoelectric}.

In this perspective, we discuss thermal transport mediated by electrons in quantum devices operating in the sub-Kelvin regime. We first present the main relevant theoretical approaches for describing charge and heat transport in nanodevices, with a particular focus on junctions formed by a gate-tunable single quantum dot (QD). Next, we discuss different experimental techniques for generating and detecting temperature gradients and heat currents. A few recent works on heat transport at the nanoscale and at ultra-low temperatures are then reviewed. Eventually, some future directions are discussed, along with broader implications of insights from thermal experiments for quantum technologies. 

\section{Theoretical Description}

In this section, we outline several theoretical approaches used to describe the flow of heat and charge through an open quantum system composed of an interacting nanostructure (QD, molecule, \mbox{etc.}) coupled to $M$ macroscopic reservoirs. The Hamiltonian of this system may be written as:
\begin{equation}
    H=H_{\rm NS} + \sum_{\alpha=1}^{M} \left[ H_{\rm R}^{(\alpha)} + H_{\rm T}^{(\alpha)} \right],
    \label{eq:H_junction}
\end{equation}
where $H_{\rm NS}$ is the Hamiltonian of the nanostructure, and each electrode is modeled as a reservoir of non-interacting Fermions such that:
\begin{equation}
    H_{\rm R}^{(\alpha)} = \sum_{k\in\alpha , \sigma}\varepsilon_{k\sigma} c_{k\sigma}^\dagger c_{k\sigma},
    %\;\;\;
    %N_\alpha=\sum_{\substack{k\in\alpha\\ \sigma}} c^\dagger_{k\sigma}c_{k\sigma},
    %%\label{eq:Hlead}
\end{equation}
where $c^\dagger_{k\sigma}$ creates a $\sigma$-spin electron of energy $\varepsilon_{k\sigma}$ in lead $\alpha$. The tunneling of electrons between the nanostructure and its electrodes $\alpha$ is described by the Hamiltonian:
\begin{equation}
    H_{\rm T}^{(\alpha)} =\sum_{k\in\alpha} \sum_{n,\sigma} \left(V_{nk}d_{ n\sigma}^\dagger c_{k\sigma}+\mathrm{H.c.}\right),
    \label{eq:Htun}
\end{equation}
where $d^\dagger_{n\sigma}$ creates an electron with spin projection $\sigma$ on the $n$th orbital of the nanostructure.  

Following \mbox{Ref.}~\onlinecite{bergfield2009thermoelectric}, the starting point for our derivation of the heat current is the fundamental equilibrium thermodynamic identity at constant volume:
\begin{equation}
    TdS = dE - \mu dN, 
\end{equation}
where $\mu$ and $T$ denote the chemical potential and the temperature, and $S$, $E$, and $N$ are entropy, internal energy, and particle number, respectively. Applying this identity to electrode $\alpha$ gives:
\begin{equation}
   \dot{Q}_\alpha \equiv T_\alpha \frac{dS_\alpha}{dt} 
    = \frac{d}{dt}\left\langle H_{\rm R}^{(\alpha)}\right\rangle - \mu_\alpha \frac{d}{dt}\left\langle N_\alpha \right\rangle,
    \label{eq:thermo_relation}
\end{equation} 
where $\dot{Q}_\alpha$ is the heat current flowing from the nanostructure into electrode $\alpha$, %is thus
and $T_\alpha$ and $\mu_\alpha$ are the temperature and chemical potential, respectively, of electrode $\alpha$.  
%and ${\displaystyle N_\alpha=\sum_{k\in\alpha,\sigma} c^\dagger_{k\sigma}c_{k\sigma}}$.
The time derivatives on the \mbox{R.H.S.} of \mbox{Eq.} (\ref{eq:thermo_relation}) may be evaluated using standard quantum mechanics to obtain:
\begin{align}
    \dot{Q}_\alpha &= \! -\frac{i}{\hbar} \left\{ \left\langle {\left[ H_{\rm R}^{(\alpha)},H \right]} \right\rangle  - {\mu _\alpha }\langle {\left[ N_\alpha,H \right]}  \rangle \right\} \nonumber \\ 
    & = \!  \frac{i}{\hbar} \! \sum_{\substack{k \in \alpha \\ n,\sigma}}  {\left( {{\varepsilon _{k\sigma }} \! - \! {\mu _\alpha }} \right)} \!\!  
    \left[ {  {V_{nk}}\langle d_{n\sigma }^\dag {c_{k\sigma }}\rangle \! - \! V_{nk}^*\langle c_{k\sigma }^\dag {d_{n\sigma }}\rangle } \right].
    \label{eq:heat_current_derivation}
\end{align}
The correlation functions in this expression may be treated using a variety of methods. We outline two commonly used theoretical frameworks below.

\subsection{Non-equilibrium Green's functions (NEGF)}

Within the non-equilibrium Green's function (NEGF) formalism \cite{evers2020advances, HaugAndJauhoBook,stefanucci2013nonequilibrium,Brandbyge02,Datta12},
%cuniberti2005introducing,Datta05,nitzan2006chemical,DiVentraBook,cuevas2010molecular,Datta12} 
the steady-state current can succinctly described as:
\begin{align}
    %I_\alpha &=& \frac{ie}{h} \int \!\! dE {\rm Tr}\left\{ \Gamma^\alpha(E)\!\left(G^<(E) + f_\alpha(E)\!\left[G(E)\!-\!G^\dagger(E)\right]\right) \right\}
    %\nonumber \\
    I_\alpha^{(\nu)}&= - \frac{i}{h} \int \!\! dE (E-\mu_\alpha)^\nu \times \nonumber \\ 
    & \!\!\!\!\!\!\!{\rm Tr}\left\{ \Gamma^\alpha(E)\!\left({\cal G}^<(E) + f_\alpha(E)\!\left[{\cal G}(E) -{\cal G}^\dagger(E)\right]\right) \right\},
    \label{eq:both_currents}
\end{align}
where $I_\alpha^{(1)}\equiv \dot{Q}_\alpha$ is the heat current \cite{bergfield2009thermoelectric} and $-eI_\alpha^{(0)}$ is the Meir-Wingreen \cite{meir1992landauer,Jauho94} expression for the charge current, $f_\alpha(E)$ %=\{1+\exp[(E-\mu_\alpha)/k_B T_\alpha]\}^{-1}$
is the Fermi-Dirac distribution for electrode $\alpha$, $\Gamma^{\alpha}(E)$ is the tunneling-width matrix describing the bonding between the nanostructure and macroscopic electrode $\alpha$, and ${\cal G}(E)$ and ${\cal G}^<(E)$ are Fourier transforms of the retarded and Keldysh ``lesser'' Green's functions:
\begin{align}
    {\cal G}_{n\sigma,m\sigma'}(t)&=-i\theta(t)\langle \{d_{n\sigma}(t),d_{m\sigma'}^\dagger(0)\}\rangle, \nonumber \\
    {\cal G}^<_{n\sigma,m\sigma'}(t)&=i\langle d_{m\sigma'}^\dagger(0)\, d_{n\sigma}(t) \rangle.
\end{align}
Both ${\cal G}(E)$ and ${\cal G}^<(E)$ originate from the time-ordered Green's functions treated on a Keldysh time-contour \cite{HaugAndJauhoBook,stefanucci2013nonequilibrium}, so any determination of ${\cal G}(E)$ also yields ${\cal G}^<(E)$ without the need for further approximation. 

Although Eq.~(\ref{eq:both_currents}) is exact, a solution to the quantum many-body problem is not generally known and approximations for the junction's Green's functions are often necessary. Typically, a system's Green's function is found through the use of Dyson's equation:
\begin{equation}
    {\cal G}(E) = \left[{\cal G}_0(E)^{-1} - \Sigma(E)\right]^{-1},
\end{equation}
where ${\cal G}$ is the system's Green's function, ${\cal G}_0$ is the ``free'' Green's function which can be solved for exactly, and $\Sigma$ is the self-energy, which describes the influence  on ${\cal G}$ of terms not included in ${\cal G}_0$. Dyson's equation is a closed form expression of a perturbation expansion taken to infinite order and gives the exact system's Green's function provided $\Sigma$ is known. Typically, however, $\Sigma$ must be approximated.

NEGF approaches may be roughly categorized into two classes: many-body perturbation theories, such as those based on the GW \cite{thygesen2007nonequilibrium,thygesen2008conserving,darancet2007ab,spataru2009gw,strange2011self} approximation to Hedin's equations \cite{hedin1965new}, the molecular Dyson equation \cite{bergfield2009many}, or the Kadanoff-Baym equations \cite{kadanoff1962quantum,dahlen2007solving,myohanen2009kadanoff}, and effective single-particle methods such as those based on current implementations of the Kohn-Sham scheme of density functional theory (KS-DFT) \cite{Brandbyge02,taylor2001ab,xue2002first,cuniberti2005introducing}. % or time-dependent DFT (TD-DFT) \cite{myohanen2008many}. 
The correlation (Green's) functions can also be treated to finite order, allowing investigations into the importance of certain correlation effects \cite{kubala2008violation,kubala2008violation}, for instance, in the thermal transport through quantum dot systems \cite{dutta2017thermal}. NEGF methods differ significantly in terms of their required complexity and computational costs, highlighting the versatility of these approaches.

\subsection{Quantum Master Equation (QME)}

The quantum master equation (QME) formalism offers an alternative approach to studying quantum transport in a nanosystem. In this method, super-operators that govern the equation of motion for the reduced density matrix are derived by tracing over the reservoir degrees of freedom \cite{harbola2006quantum,Harbola08,PhysRevB.74.235309,konig1997cotunneling,pedersen2005tunneling,leijnse2008kinetic,timm2008tunneling}. The tunneling Hamiltonian is treated perturbatively within the QME formalism, which restricts its application to systems with weak couplings between the nanostructure and reservoirs (e.g., $\langle \Gamma^\alpha \rangle \ll kT$). However, once the (approximate) density matrix density is determined, all observable quantities can be found directly via von Neumann's equation, making the theory ideal for investigating time-dependent properties.

In cases where coherence between states can be disregarded, transport can be described by a set of intuitive rate equations \cite{Muralidharan06,beenakker1991theory,bonet2002solving,gurvitz1998rate,PhysRevB.47.4603,pedersen2005tunneling}. These rate equations offer a simplified representation of the transport processes but often capture salient mechanisms \cite{beenakker1991theory,bonet2002solving,pedersen2005tunneling}. The combination of QME formalism and rate equations provides a comprehensive framework for studying the transport properties of nanoscale systems in various scenarios.

\subsection{Elastic cotunneling regime}
	
In many cases of interest, %in nanostructures, 
elastic processes dominate the transport and Eq.~(\ref{eq:both_currents}) can be simplified giving the Landauer-B\"uttiker expression for the current \cite{Buttiker86,sivan1986multichannel}:
\begin{equation}
    \label{eq:Buttiker}
    I^{(\nu)}_\alpha=\frac{1}{h} \sum_{\beta=1}^M\int dE\; (E-\mu_\alpha)^\nu \, {T}_{\alpha\beta}(E)\left[f_\beta(E)-f_\alpha(E)\right],
\end{equation}
where $\alpha$ and  $\beta$ label the electrodes and 
\begin{equation}
    {T}_{\alpha\beta}(E)={\rm Tr}\left\{ \Gamma^\alpha(E) {\cal G}(E) \Gamma^\beta(E) {\cal G}^\dagger(E)\right\}
    \label{eq:transmission}
\end{equation}
is the transmission function with the tunneling-width for lead $\alpha$ being given by:
\begin{equation}
    \Gamma^\alpha_{nm}(E) = 2\pi \sum_{k \in \alpha} V^{\phantom{}}_{nk} V^*_{mk} \delta\left(E-\varepsilon_{k}\right).
\end{equation}
Although Eq.~(\ref{eq:Buttiker}) resembles the non-interacting result \cite{MahanBook}, it follows directly from Eq.~(\ref{eq:both_currents}) and is valid even for strongly interacting systems, provided the inelastic contribution to transport is negligible. Cotunelling processes, \mbox{i.e.} charge transfer involving intermediate (virtual) states of the nanostructure, are naturally included in this formalism, where the total transmission is a coherent mixture of the Green's functions which include contributions from all states via Dyson's equation.

In linear response, i.e. when $|\Delta \mu| \ll \mu$ and $\Delta T \ll T$, we can write %$f_\beta - f_\alpha \approx (-\partial f_0/\partial E)\Delta E$,
$f_\beta - f_\alpha \approx (-\partial f_0/\partial E) [\Delta \mu + (E-\mu)\Delta T/T]$,
%\begin{equation}
%	f_\alpha(E)\cong f_0(E)+\left(- \frac{\partial f_0}{\partial E} \right) \left[ \Delta\mu_\alpha + \frac{E-\mu}{T}\Delta T_\alpha \right],
%\end{equation} 
where $\Delta \mu =\mu_\alpha - \mu_\beta$, $\Delta T = T_\alpha - T_\beta$, and $f_0(E)$ is the equilibrium (\mbox{i.e.} zero-bias) Fermi distribution with chemical potential $\mu$ and temperature $T$. In this regime, Eq.~(\ref{eq:Buttiker}) may expressed as:
\begin{equation}
	\left( \begin{array}{*{20}{c}}
			{I_\alpha}  \\ %		{I^{\rm Q}_\alpha}  \\
   		\dot{Q}_{\alpha} \\
	\end{array} \right) = %\frac{1}{h} 
	\sum\limits_\beta  {\left( {\begin{array}{*{20}{c}}
				{ {\cal L}_{\alpha\beta}^{\left( 0 \right)}} & {\frac{1}{T} {\cal L}_{ \alpha\beta }^{\left( 1 \right)}} \vspace*{1mm} \\
				{{\cal L}_{ \alpha\beta }^{\left( 1 \right)}} & {\frac{1}{T}{\cal L}_{ \alpha\beta  }^{\left( 2 \right)}}  \\
		\end{array}} \right)\left( {\begin{array}{*{20}{c}}
				{{V_\beta } - {V_\alpha }} \vspace*{2mm} \\
				{{T_\beta } - {T_\alpha }}  \\
		\end{array}} \right)},
	\label{eq:current_matrix}
\end{equation}
where $V_\alpha = -\mu_\alpha/e$, $I_\alpha = -e I_{\alpha}^{(0)}$ is the charge current, $\dot{Q}_\alpha$ is the heat current, and the Onsager linear-response functions are given by:
%-$e$ is the electron charge, $I_\alpha\equiv -eI_\alpha^N$ is the electrical current and 
\begin{equation}
	{\cal L}^{(\nu)}_{\alpha\beta }\left(\mu,T\right) = 
	\frac{1}{h} \int dE (E-\mu)^{\nu} \left(-\frac{\partial f_0}{\partial E}\right) \,{ T}_{\alpha \beta}(E).
	\label{eq:L_function}
\end{equation}
With these functions we can compactly encode a number of important transport properties:
\begin{align}
	\label{eq:G}
	G_{\alpha\beta} &= \frac{dI}{dV} = e^2{\cal L}^{(0)}_{\alpha\beta}, \\
  \label{eq:S}
 S_{\alpha\beta} &= - \left.\frac{\Delta V}{\Delta T}\right|_{I=0} = -\frac{1}{eT} \frac{{\cal L}^{(1)}_{\alpha\beta}}{{\cal L}^{(0)}_{\alpha\beta}}, \\
	\kappa_{\alpha\beta}  &= \left.\frac{dQ}{d(\Delta T)}\right|_{I=0} = \frac{1}{T}\left({\mathcal L}^{(2)} -\frac{\left[{\cal L}^{(1)}_{\alpha\beta} \right]^2}{{\cal L}^{(0)}_{\alpha\beta}} \right),
	\label{eq:kappa}
\end{align}
where $G$ is the electrical conductance, $S$ is the thermopower, and $\kappa$ is the electronic contribution to the thermal conductance. The linear-response transport coefficients of an interacting system thus have a structure identical to that of a non-interacting system, except that ${ T}_{\alpha \beta}(E)$ must be calculated using the interacting Green's functions.  

The transport theories outlined here provide a comprehensive framework to investigate the heat, charge and spin transport through quantum systems. The influence of electron–electron correlations, coherent wave-like effects, multiple electrodes \cite{butcher1990thermal}, molecular vibrations, photo emission, \mbox{etc.} can all be described (in principle) exactly. This paves the way for exploring a variety of interesting fundamental questions in quantum thermodynamics \cite{alicki2018introduction,myers2022quantum}, including probing the applicability of the laws of thermodynamics to open quantum systems and nonequilibrium conditions \cite{shastry2019third,de2020unraveling}, as well as understanding Landauer's principle \cite{landauer1961irreversibility}, a foundational concept connecting thermodynamics and information theory.
The quantum theory of temperature measurement, first pioneered by Engquist and Anderson \cite{engquist1981definition}, can also be investigated for systems operating both in and out of equilibrium \cite{zhang2019local,dubi2011colloquium,bergfield2013probing,meair2014local,sanchez2011thermoelectric,hajiloo2020quantifying}.

\subsection{Quantum dot models}

We focus on the transport through systems composed of a small metallic or semiconductor QDs tunnel coupled to source and drain electrodes and capacitively coupled to a third gate electrode. Typical quantum dots have between $10^3-10^9$ atoms with an equivalent range of electrons. Current nanofabrication techniques allow the size and shape, and therefore the electronic properties of these {\em artificial atoms} to be precisely engineered.

At low temperature, and small bias voltage, the energy to add an electron onto a QD, $E_C=e^2/2C$, can exceed the thermal energy $kT$ and the total tunneling coupling energy $\hbar(\Gamma^S + \Gamma^D)$, where $C$ is the self-capacitance of the QD, and $\Gamma^S$ and $\Gamma^D$ are the tunneling coupling to the source and drain electrodes, respectively, an effect known as Coulomb blockade \cite{kouwenhoven1997electron}. In this regime, which for typical QDs with capacitance values on the order of femto-Farads corresponds to sub-Kelvin experiments, the granular nature of charge can be directly observed from electron transport, for instance, via the observation of the single-electron transistor (SET) effect where the conductance exhibits strong variations (Coulomb oscillations) as a function of gate voltage.

In many QD systems, transport may be accurately described by a single-level model with the following Hamiltonian 
\begin{equation}
%H_{QD} = \sum_{\sigma}\varepsilon_{\sigma} n_\sigma + Un_\uparrow n_\downarrow
H_{QD} = \varepsilon \left (n_\uparrow + n_\downarrow \right) + Un_\uparrow n_\downarrow
\end{equation}	
where $n_\sigma = d^\dagger_\sigma d_\sigma$ is the number operator and $U$ is the Coulomb energy.  

The eigenstates of this model, often called an Anderson model \cite{PhysRev.124.41} can be solved exactly using Bethe Ansatz techniques \cite{wiegmann1983exact}. At zero temperature the linear-response transmission between the source and drain electrodes is given by \cite{pustilnik2004kondo,PhysRevLett.87.236801}
\begin{equation}
    T(E) = \frac{4 \Gamma^S \Gamma^D}{(\Gamma^S + \Gamma^D)^2} \sin^2 \left[\frac{\theta(E)}{2} \right]
\end{equation}
where the total number of electrons on the central region $\langle n_C \rangle$ is related to the eigenphases $\theta(E)$ by the Friedel-sum rule \cite{PhysRev.150.516,friedel1958metallic}
\begin{equation}
    \langle n_C \rangle = \theta(E) / \pi.
\end{equation}
In the case where dynamic electron-electron correlations can be neglected (\mbox{i.e.} when the $\Sigma(E) \equiv \Sigma$), the transmission function may be expressed as a single Lorentzian
\begin{equation}
    {T} (E)= %\frac{4 \Gamma^L\Gamma^R}{(\Gamma^L + \Gamma^R)^2}
    \frac{ 4\Gamma^L \Gamma^R}{(E- \tilde{\varepsilon})^2 + (\Gamma^L + \Gamma^R)^2},
    \label{Eq:transQD}
    \end{equation}
where $\tilde{\varepsilon} = \varepsilon - e\alpha V_g$ is the effective on-site potential shifted by the applied gate voltage $V_{\rm g}$. With the transmission function one can then evaluate the thermoelectric coefficients in linear response using Eqs.~(\ref{eq:L_function}-\ref{eq:kappa}).

\section{Experimental techniques}

In this section, we discuss different experimental approaches for thermoelectric and heat conductance measurements in quantum devices. All of these invariably require establishing a thermal imbalance over nano- or micrometer distances. This implies that both the heating and the measurement of temperatures must be local at this scale. We limit ourselves to the discussion of purely electronic experiments, which excludes for instance far-field thermal mapping \cite{ward2011vibrational,halbertal2016nanoscale} or phonon heat conductance experiments \cite{schwab2000measurement,heron2009mesoscopic,zen2014engineering}. Although some experiments have demonstrated heat flow studies in response to local on-chip {\it cooling} \cite{edwards1995cryogenic,pekola2000microrefrigeration,meschke2006single,chowdhury2009chip,pascal2013existence,dutta2017thermal}, the most common approach consists in locally overheating the sample or chip, by Joule dissipation. 

For measurements of the Seebeck coefficient $S$, %=-\Delta V/\Delta T \mid_{I=0}$, 
the experimental quantity of interest is a voltage in response to a temperature gradient. Such experiments do not require knowledge of the heat currents involved. If only the variation of $S$ with respect to other control parameters, such as magnetic field or gate, is investigated, it may not even be necessary to measure $\Delta T$, as long as it can be considered constant \cite{mitdank2012enhanced,moon2013gate,small2003modulation,dutta2018direct}. The same holds for non-quantitative thermoelectric coefficient measurements.

On the other hand, heat conductance and {\it quantitative} thermopower measurements require that $\Delta T$ is known. This implies introducing local thermometers. There are actually a limited number of experiments in which this is actually done \cite{dutta2017thermal,maillet2020electric,dutta2020single,majidi2022quantum,sivre2018heat,jezouin2013quantum}, while others deduce the thermal gradients indirectly, based on models \cite{van1992thermo,josefsson2018quantum,dutta2018direct,roddaro2014large,hoffmann2009measuring}. Eventually, measurements of heat conductance $\kappa={\dot Q}/\Delta T$ further require knowledge of the heat flow ${\dot Q}$, which is the most challenging quantity to determine quantitatively. This section is thus divided into two parts, discussing the determination of $\Delta T$ and ${\dot Q}$, respectively, across a quantum device.

\subsection{Local electron thermometry}

Because electron-phonon or electromagnetically mediated heat flows are very small at millikelvin temperatures \cite{pascal2013existence}, it is crucial to directly access the local {\it electronic} temperature on each side of the device. For this, the most common approach is to measure an electrical quantity depending on the electronic temperature. To start with, one must ensure that a temperature can indeed be defined in the lead, which may not necessarily be the case in quantum conductors brought out-of-equilibrium by a strong enough current bias for instance \cite{pothier1997energy}. 

\subsubsection{Thermometry based on a resistor}
A simple strategy to measure a local electronic temperature $T$ is to use an electrode material with a temperature-dependent bulk resistivity $\rho(T)$. At low temperatures, materials exhibiting a non-saturating resistivity are mostly either quite disordered or have a low electronic density, and are at the transition to the insulating state. Some examples include nanowires of NbN \cite{nguyen2019niobium} or Pt-C \cite{blagg2020focused}, as well as highly doped Si \cite{adami2018characterization}, which is used in microscale bolometers. Because of their usually large resistance, read-out related overheating can be an issue.

A thermometry method, which can under proper circumstances even provide a primary temperature reading, is taking advantage of the thermal (Johnson-Nyquist) noise of an unbiased resistance $R$, leading to voltage fluctuations with a mean square amplitude $\langle v_T^2 \rangle=4k_BTR \,\Delta f$, where $k_B$ is Boltzmann’s constant and $\Delta f$ is the measurement bandwidth \cite{PhysRev.32.97}. In experiments on quantum point contacts (QPCs) in two-dimensional electron gases, the thermal noise in a quantum point contact, tuned to exactly one quantum of electrical conductance, was used as an integrated thermometer of a nearby overheated metallic structure, from which the heat conductance through other channels connected to the same island is then inferred \cite{jezouin2013quantum}. In practice, the voltage fluctuations must be measured up to  MHz frequencies, which requires embedding the noisy resistor into a matched resonant circuit, in combination with a cryogenic noise amplification scheme \cite{talanov2021high}. Remarkably, local electronic noise thermometry in a quantum device was demonstrated without saturation down to 7 mK \cite{iftikhar2016primary}. One drawback is related to the long necessary integration times, due to the smallness of thermal fluctuations in the lower millikelvin range, which make applications for time-resolved thermometry challenging.

\subsubsection{Thermometry based on hybrid superconducting junctions}
\begin{figure*}[t!]
	\includegraphics[width=4.0in]{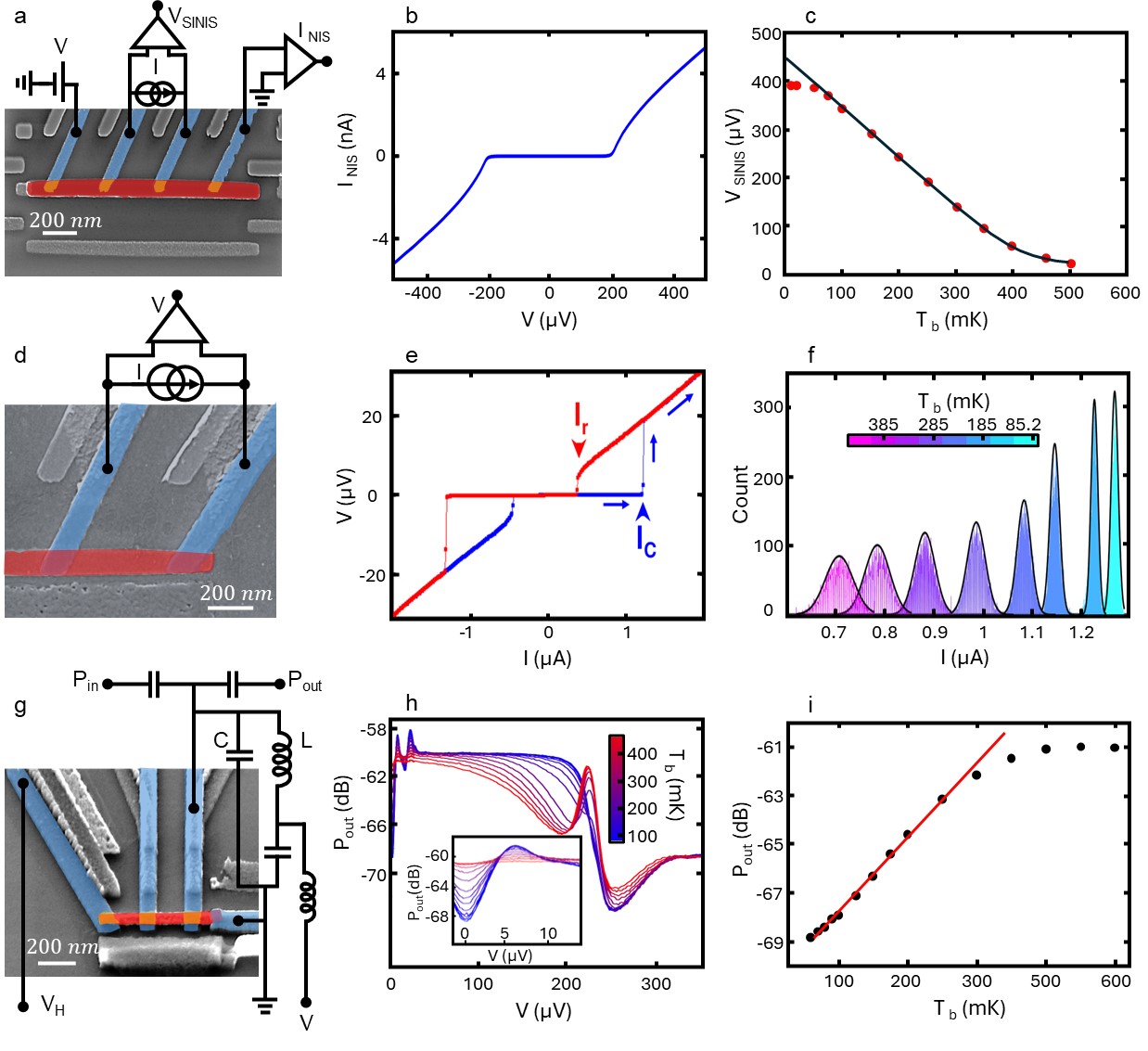}
	%\centering
 \vspace{-0.4cm}
	 \caption{Superconducting hybrid electron thermometers. a) Scanning electron micrograph (SEM) of a NIS junctions with N, S and the aluminum oxide tunnel junctions (I) highlighted respectively in red, blue and orange. Five superconducting leads allow for both heating and thermometry. b) Current–voltage ($IV$) characteristics of an NIS junction at a bath temperature $T_b= 100$ mK ($R_T= 85.6$ k$\Omega$). c) Voltage drop across the NIS junction with a fixed $I= 5$ pA (highlighted in b) as a function of $T_b$. The red line represents the expected response from \mbox{Eq.} (\ref{eq-I-NIS}) with gap $\Delta=$ 209 $\mu$eV and tunnel resistance $R_T= 96.2$ k$\Omega$. d) SEM of an SNS junction (same color code as above, except that the contacts are now transparent, which is symbolised by purple color), probed in a current-biased configuration. e) $IV$ characteristics of the SNS thermometer junction at $T_b$ = 75 mK, featuring a critical (switching) current $I_c$ and retrapping current $I_r$. f) Histograms of the stochastic switching current distribution at various bath temperatures with a gaussian envelope. g) Device schematics with the same color code as above. Note the coexistence of the two types of junctions. The SINS junction is embedded into an RF tank circuit. h) RF transmitted power $P_{out}$ at resonance of the SINS junction within the tank circuit as a function of bias $V$ applied to the junction (only $V\ge 0$ shown), at different cryostat temperatures. The inset highlights the strong temperature dependence of the zero-bias Josephson conductance. i) Calibration of $P_{out}$($V=$0) against cryostat temperature at equilibrium with a linear fit excluding points beyond $T_b >$ 250 mK. b,c) Adapted with permission from \mbox{Ref.} (\onlinecite{majidi2022quantum}). Copyright 2022 American Chemical Society. a,d,e) Adapted with permission from \mbox{Ref.} (\onlinecite{majidi2021heat}). f) Adapted with permission from \mbox{Ref.} (\onlinecite{dutta2020single}). Copyright 2022 American Physical Society.}
	\label{HybridThermometry}
\end{figure*}

When a thin insulator I makes the junction between a superconductor S and a normal metal N, the conductance is strongly suppressed at low bias voltages $|V|<\Delta/e$, with $\Delta$ the gap of the single-particle excitation spectrum of S. The current flowing can be expressed as:
\begin{equation}
\label{eq-I-NIS}
    I=\frac{1}{2eR_N}\int dE \,(f_N(E-eV)-f_N(E+eV))\rho_S(E),
\end{equation}
with $\rho_S$ is the normalized electronic density of states of the superconductor. As $|V|$ approaches $\Delta/e$, a thermally activated quasiparticle current is allowed, making NIS junctions very sensitive to the electron temperature in N \cite{Nahum93}, as shown in \mbox{Fig.} \ref{HybridThermometry}a-c. Except for the temperature dependence of $\Delta$, which can be neglected below about $T_c/2$ (with $T_c$ the superconducting transition temperature), NIS junctions are totally insensitive to the temperature in the S contact. Due to their high responsivity and relatively easy integration with other metallic circuits, NIS junctions are widespread millikelvin local electron thermometers \cite{dutta2017thermal,majidi2022quantum}. In practice, two NIS junctions are often used in series, via a common central N island, in the co-called SINIS geometry.

One crucial drawback of NIS (or SINIS) thermometers is that the rather large operating voltage $\approx \Delta/e$ inherently lead to significant dissipation in S (see next subsection for a detailed discussion of the heat balance). Part of this heat may return to N by various processes, leading to a saturation of the sensitivity near 50 mK. An alternative approach consists in suppressing the tunnel barriers in SINIS junctions. In this case, superconducting correlations may coherently propagate from one S lead to the other, allowing for a finite dissipationless Josephson current \cite{dubos2001josephson}. In such typically sub-micrometer-long SNS junctions, the relevant energy scale is no longer $\Delta$, but the Thouless energy $E_{\rm Th}$, related to the electronic diffusion time over the length $L$ of the normal metallic island N. The amplitude of the Josephson critical current $I_c$ is affected by depairing in N via the dimensionless parameter $E_{\rm Th}/k_BT$, making it a possible thermometer (see \mbox{Fig.} \ref{HybridThermometry}d-f). However, probing $I_c$ in SNS junctions implies a transition to the dissipative state. The associated thermal runaway was shown to be at the origin of the hysteresis generally found in these devices \cite{courtois2008origin}. Therefore, $I_c$ measurements are not reversible and bias must be continuously reset. For temperatures below $E_{\rm Th}/k_B$, $I_c(T)$ saturates, suggesting the use of small-$E_{\rm Th}$ (and thus small $I_c$) junctions when going to lower temperatures. This however makes the voltage jump at $I_c$ harder to detect. In practice, due to the above-described effects, SNS thermometry is also losing sensitivity below about 50 mK.

Recently, a superconducting hybrid device based on a combination of a transparent and a weak/semi-transparent tunnel junction (noted i, in contrast with the more opaque barrier I) has been demonstrated to be less prone to saturation at low temperatures \cite{karimi2018noninvasive}. In such an SINS structure, the transparent S contact is used to induce superconducting correlations in N. A nearby semi-transparent tunnel junction to a second S contact is thus effectively connecting an intrinsic and a proximized superconductor. The Josephson correlations across this tunnel junction give rise to a conductance peak at zero bias voltage, which is highly sensitive to the temperature in N (see \mbox{Fig.} \ref{HybridThermometry}g-i). Such a device avoids the shortcomings of both previously discussed superconducting hybrids: it is reversible and can be operated near zero voltage, implying much lower dissipation levels. It was shown to operate down to 25 mK without saturation \cite{karimi2020reaching}. 

\subsubsection{Thermometry based on quantum capacitance}
Finally, we discuss thermometry applications of quantum dot (QD) junctions, in which a single quantum level (due to quantum confinement) is tunnel coupled to a single or two macroscopic contacts (Fermi seas) \cite{torresani_nongalvanic_2013,ahmed_primary_2018-1,nicoli_quantum_2019-1,chawner_nongalvanic_2021-1}. 

The occupation of a QD level coupled to one lead only depends on its relative position $\epsilon$ with respect to the contact's Fermi level $\mu$ and temperature. The gate-dependent quantum capacitance of a QD level fluctuating between occupation states 0 and 1 may be expressed as:
\begin{equation}
\label{eq-QD}
    C(V_g)=\frac{\alpha^2}{4k_BT}\cosh^{-2}\left(\frac{\alpha(V_g-V_g^0)}{2k_BT}\right),
\end{equation}
with $V_g$ the gate voltage acting on the QD level, $\alpha =\partial \epsilon/\partial V_g$ the gate lever arm, and $V_g^0$ the gate voltage at charge degeneracy, at which the empty and singly occupied states are equally probable. Note that for a spin-degenerate level with a charge state fluctuating between 0 and 1, charge degeneracy is not occurring exactly at $\epsilon=\mu$, but at an energy $k_BT\ln(2)$ below. This implies that the charge degeneracy point itself is slightly temperature dependent. Both the magnitude of $C(V_g^0)\propto 1/T$ at charge degeneracy and the full-width at half maximum of $C(V_g)$, equal to $\alpha\Delta V_g=4\ln(\sqrt{2} +1)k_BT\approx3.5 k_BT$, can be used for thermometry purposes. The above description is only accurate in the weak coupling regime, that is, as long as the thermal energy $k_BT$ largely exceeds the tunnel coupling $\hbar\Gamma$. At low temperatures, $\alpha \Delta V_g$ saturates to $2\hbar \Gamma$.

\subsubsection{High-frequency thermometry}
We conclude this panorama of local electron thermometry techniques in quantum devices with a discussion of recent developments in time-resolved readout methods, allowing for measurements with a MHz bandwidth. To this end, either the tunneling resistance or the capacitance used for thermometry is embedded into a radio-frequency resonant circuit \cite{gasparinetti2015}. The change of transmitted or reflected power of the resonator then allows for a fast read-out of changes in the temperature-dependent sensor impedance, as in the experiments shown in \mbox{Fig. \ref{HybridThermometry}g-i}. The readout bandwidth is then essentially set by the resonator bandwidth, which can be adjusted to exceed several MHz. This has allowed the determination of the electron-phonon energy relaxation rates in a variety of nanoscale conductors \cite{gasparinetti2015, viisanen2018, karimi2020reaching}, as well as detecting and quantifying minute dissipative events such as individual phase slips in a Josephson junction \cite{gumus2023}. Time-resolved experiments thereby open formidable perspectives for investigating fluctuations, stochasticity and quantum effects in thermodynamics \cite{pekola2022,champain2023real}. 

\subsection{Heat balance}

\begin{figure}[t!]
	\includegraphics[width=\columnwidth]{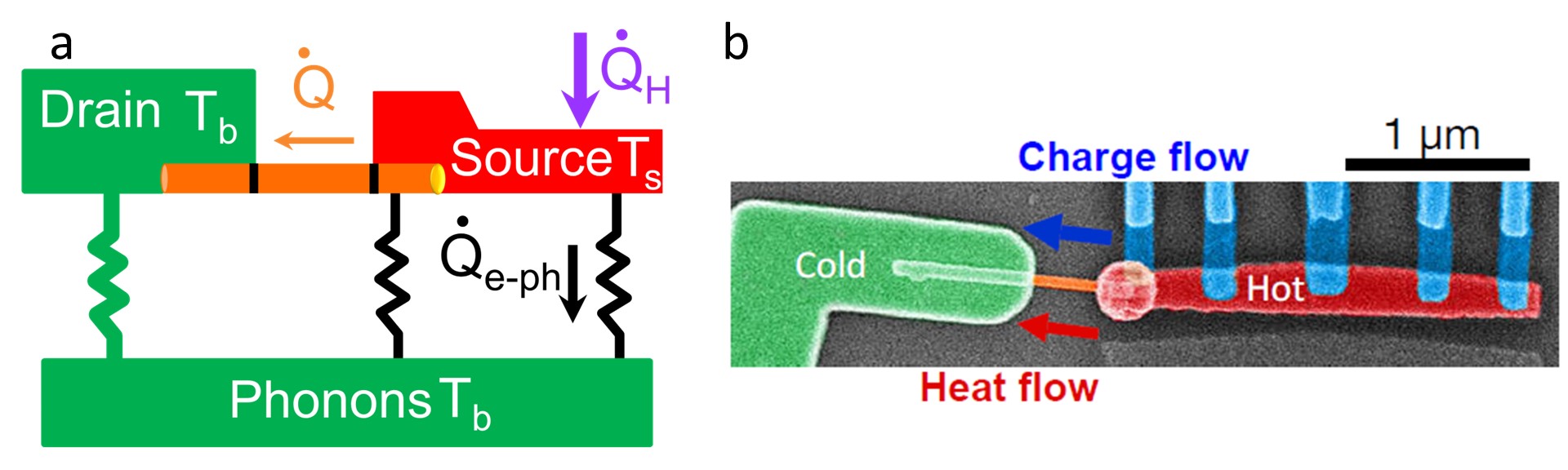}
	%\centering
 \vspace{-0.4cm}
	\caption{(a) Thermal balance diagram, highlighting the heat load $\dot Q_H$ to the thermally floating source, the parasitic heat leakage, here in the form of $\dot Q_{\rm e-ph}$, and the heat flow to the cold drain across the device under study, here a nanowire. (b) False colored scanning electron micrograph of the bolometer for measuring heat transport across an InAs nanowire (orange). The superconducting leads, forming NIS tunnel junctions to the Cu normal island are shown in cyan. The color code is in correspondence with (a). Adapted with permission from Ref. (\onlinecite{majidi2022quantum}). Copyright 2022 American Chemical Society.}
	\label{schematics}
\end{figure}

The typical experimental geometry used for the results presented here is shown in Fig. \ref{schematics}. Alike most heat conductance experiments, it is designed such that one contact to the device is well thermalized to the environment, usually via low-resistance (but not superconducting) electrical conduction to macroscopic contacts. In this part of the device, electrons and phonons are in equilibrium and are part of a large thermal bath at temperature $T_{\rm bath}$ at all times, thus acting as a thermal {\it drain} at temperature $T_d$, such that $T_d=T_{\rm bath}$. If this is granted, a single temperature measurement in the thermally floating part of the device ({\it source} in the remainder) at temperature $T_s$ is sufficient for determining $\Delta T=T_s-T_d$ across the device, as exemplified in Fig. \ref{Nw+heatvalve}a,b.

A thermal transport measurement across a nanoscale conductor requires having a quantitative knowledge of the heat flow $\dot Q$ in response to a given temperature landscape. In linear response, one considers only small temperature gradients $\Delta T\ll T$, which leads to defining the electronic thermal conductance $\kappa={\dot Q}/\Delta T$, in line with the Onsager relation \mbox{Eq.} (\ref{eq:kappa}). However, it is often useful to go beyond linear response and determine the full $\dot Q(T_d,T_s)$ curve, for a wide range of drain and source temperatures $T_d$ and $T_s$. On an important side note, heat currents and thus $\kappa$ are defined at zero average particle current, and thus in open circuit conditions. 

\subsubsection{Electronic heating}
In practice, a known heating power $\dot Q_H$ is applied to the source, part of which is then flowing through the conductor under investigation, while another part $\dot Q_s$ is escaping the source by parasitic paths, such that $\dot Q= \dot Q_H - \dot Q_s$. The heating power $\dot Q_H$ is usually provided by similar architectures than used for thermometry in the same device. In the case of ohmic elements, such as for a bulk resistive electrode or a quantum Hall channel, $\dot Q_H$ is easily estimated from Joule's law. In the case of an NIS tunnel junction, one must start from the full heat transport equation to determine the power applied to N, yielding:
\begin{equation}
\label{eq-Q-NIS}
    \dot Q_H^{N}=\frac{1}{e^2R_N}\int dE \, (E-eV)(f_N(E-eV)-f_S(E))\rho_S(E).
\end{equation}
Because of the strong energy dependence of $\rho_S$ near $\pm\Delta$, $\dot Q_H^{N}$ is strongly non-linear in the applied voltage $V$, and can even be negative for $|V|<\Delta/e$, leading to cooling in N \cite{giazotto2006opportunities}. This peculiar feature allowed testing the response of heat flows on inverting the sign of $\Delta T$ \cite{pascal2013existence,dutta2017thermal}. The heating power applied to the superconducting contact is obtained by replacing $\mu_N$ by $\mu_S$ in \mbox{Eq.} (\ref{eq-Q-NIS}), which leads to $\dot Q_H^S=IV-\dot Q_H^N$. This quantity is always positive and in can be quite large, such that a fraction of $\dot Q_H^S$ can spuriously return to N via other mechanisms, such as transport through phonons. It was found that a good practical description of the heat generation in N near an NIS junction writes $\dot Q_H=\dot Q_H^{N}+\alpha \dot Q_H^S$, with $\alpha$ on the order of a few percent \cite{ullom2000,pascal2013existence}, which however depends on the geometry of NIS junction and in particular of the S lead.

Although the heating power delivered by an NIS junction is not as trivial as from a simple ohmic element, it still remains under good control, and even the contribution of higher order (Andreev) tunneling processes can be accounted for \cite{rajauria2008andreev}.

\subsubsection{Optimization of thermal isolation}
One important source of heat leakage is associated to electronic conduction through the leads, which are unavoidable for biasing the device and creating/measuring the temperature gradients. High-resistance (tunnel) contacts are obviously preferable for confining heat. In the particular case of NIS junctions, the strong conductance suppression at low voltages provides a very efficient isolation. However, it is usually desirable to enable at least one low-resistance access to the source, in order to be able to characterize the electrical properties of the device under investigation. In the case of experiments performed using edge channels in two-dimensional electron gases, the gate-tunability of the different contact resistances is of course a great advantage. In the other case of superconducting contacts to a normal metallic source, be it via opaque or transparent contacts, the thermal isolation can strongly benefit from the exponentially suppressed heat conductance of superconductors at very low temperature \cite{peltonen2010thermal}. Because of the inverse proximity effect at a NS interface, which appears particularly at high contact transparencies, the superconducting order parameter is however suppressed near the interface over a distance $\sim \xi$, the superconducting coherence length. It is thus crucial that the superconducting thermal isolation be conceived with a length much longer than $\xi$ until the next effective heat drain (wider or normal conducting electrodes).

With well-designed superconducting leads (furthermore if connected by tunnel contacts), electronic conduction is not the dominant heat escape channel at very low temperature. In practice, and assuming aluminum as the superconducting material, the combination of heat leakage at high temperatures $T>T_c/4$ and thermometry saturation at low temperatures unfortunately does not permit NIS-thermometer/heater-based bolometers to perform quantitative electronic heat conductance measurements over a temperature range exceeding roughly a decade, between about 30 and 300 mK at best. A wider accessible temperature window would be highly desirable, for instance for investigating Kondo physics in quantum dot junctions \cite{costi2010thermoelectric,dutta2018direct}, but is presently challenging to achieve \cite{iftikhar2018}.

\subsubsection{Electron-phonon coupling}
Assuming that the spurious heat leakage via the contacts is negligible, the dominant origin of $\dot Q_s$ is in general the electron-phonon coupling in the source electrode, which takes the form $\dot Q_{\rm e-ph}={\mathcal V}\Sigma \left( T^n-T_{\rm ph}^n \right)$. Here $\mathcal V$ is the interaction volume, $\Sigma$ the material-dependent electron-phonon coupling constant, $n$ a number ranging between 4 and 6 depending on the material's phonon properties, and $T_{\rm ph}$ the phonon temperatures. The fast decay of the electron-phonon coupling in the millikelvin regime is the key ingredient allowing for thermally isolating small regions from the environment, without the need for instance for suspended structures. To put some numbers, at a 100 mK bath temperature, the electrons in a $\mathcal V = 2\times0.2\times0.05\,\mu$m$^3$ copper bolometer ($n=5$, $\Sigma=2\times 10^{-9}$ W.$\mu$m$^{-3}.K^{-5}$) overheated by $\Delta T=T- T_{\rm ph}=10$ mK emit a power $\dot Q_{\rm e-ph}=200$ aW to the phonon bath. This is to be compared to the heat current across a single spin-degenerate quantum conduction channel under the same temperature bias and obeying the Wiedemann-Franz law, $\dot Q \approx L_0\,\Delta T \,T\, G_0=1.85$ fW. When for instance investigating the heat flow in one or a few quantum conduction channels, obeying order-of-magnitude-wise the WF law, while working at temperatures below 30 mK, is appears that $\dot Q_{\rm e-ph} \ll \dot Q$ and spurious heat leakage of any kind can be neglected. This was the case in several experiments in two-dimensional electron gases under high magnetic fields \cite{jezouin2013quantum,sivre2018heat}.

On the other hand, the combined use of NIS-type bolometers, which do not allow operating in the low mK temperature range, together with the investigation of quantum dot junctions, which have typical charge conductances that are only a fraction of $G_0$, make that $\dot Q_{\rm e-ph}$ and $\dot Q$ have the same order of magnitude \cite{dutta2020single}. For quantitative measurements, it is thus necessary to be able to estimate $\dot Q_{\rm e-ph}$ accurately. One possibility is of course to predict $\dot Q_{\rm e-ph}$ from literature values for a given bolometer material. This is however not very satisfactory, as the strength of the electron-phonon coupling shows a certain variability, depending on the sample geometry and material preparation procedure. 

\subsubsection{Independent determination of the heat balance parameters}
When $\dot Q$ can be tuned by some external parameter $\lambda$, such as a gate voltage, there can be one or more peculiar values $\lambda_0$ at which $\dot Q$ can be predicted with high confidence. For instance, in a quantum dot junction, tuning the device in gate-space far away from a conduction resonance allows assuming that $\dot Q$, which is by initial hypothesis associated to electron transport through the dot, is also nil. When measuring the full $\dot Q_H(\lambda_0, T_s, T_{\rm bath})$ curve, this experimental quantity can be identified with $\dot Q_s$, that is, usually $\dot Q_{\rm e-ph}$ (see Fig. \ref{Nw+heatvalve}c). In practice $\dot Q_H$ is swept and $T_s$ is measured, but the additivity of powers and the bijectivity of the relation makes it preferable and licit to represent it the other way around) at a fixed $T_{\rm bath}$. Making the assumption that $\dot Q_s$ does not depend on $\lambda$, which should be valid if $\lambda$ acts only on the device under investigation and not on the bolometer itself, one can then write for any $\lambda$ that \cite{majidi2022quantum}:
\begin{equation}
\label{eq-Q-balance}
\dot Q(\lambda,T_s,T_{\rm bath})= \dot Q_H(\lambda, T_s, T_{\rm bath})-\dot Q_H(\lambda_0, T_s, T_{\rm bath}).
\end{equation}

It is of course important to verify {\it a posteriori} that the form of $\dot Q_H(\lambda_0, T_s, T_{\rm bath})$ can be reasonably described using an electron-phonon heat escape model, or anything equivalent. For instance in \mbox{Ref.} \cite{majidi2022quantum}, which reported heat transport measurements through a quantum dot formed in an InAs nanowire, about half of the nanowire was part of the bolometer itself. It was observed that $\dot Q_s$ itself displayed at small and smooth but measurable gate dependence, which was attributed to an enhanced electron-phonon coupling in the nanowire itself as the gate voltage, and thus the charge carrier density, was cranked up. This issue could however be sorted by performing the heat balance subtraction in Eq.\ (\ref{eq-Q-balance}) locally, for $\lambda\sim\lambda_0$. In the analysis of another heat transport experiment, through a single-electron transistor (SET) \cite{dutta2017thermal}, such non-conducting points in gate space did not always exist, due to high junction transparencies. Therefore it was assumed that the heat conductance at the charge degeneracy points did obey the WF law, as theoretically predicted \cite{kubala2008violation}. This then allowed deducing $\dot Q_s$ from the heat balance at the charge degeneracy point, and subtracting it to obtain $\dot Q$ at all gate voltages. 

\section{Quantum signatures in heat transport experiments}

\subsection{Electronic thermal conductance}

Deviations from the WF law for the thermal conductance indicate a breakdown of the free-electron model and can act as a signature of quantum effects. This discussion will address two primary sources of such deviations: energy-dependent transmission coefficients $T(E)$ due to quantum confinement and interactions. Other mechanisms exist, such as the unique quantum statistics in two-dimensional electron gases \cite{banerjee2018,dutta2022isolated,srivastav2022} or quantum wave effects \cite{bergfield2009thermoelectric}, where the difference in influence of quantum interference on the charge and heat transport generate significant violations of the WF law.

 \begin{figure}[t!]
	\includegraphics[width=\columnwidth]{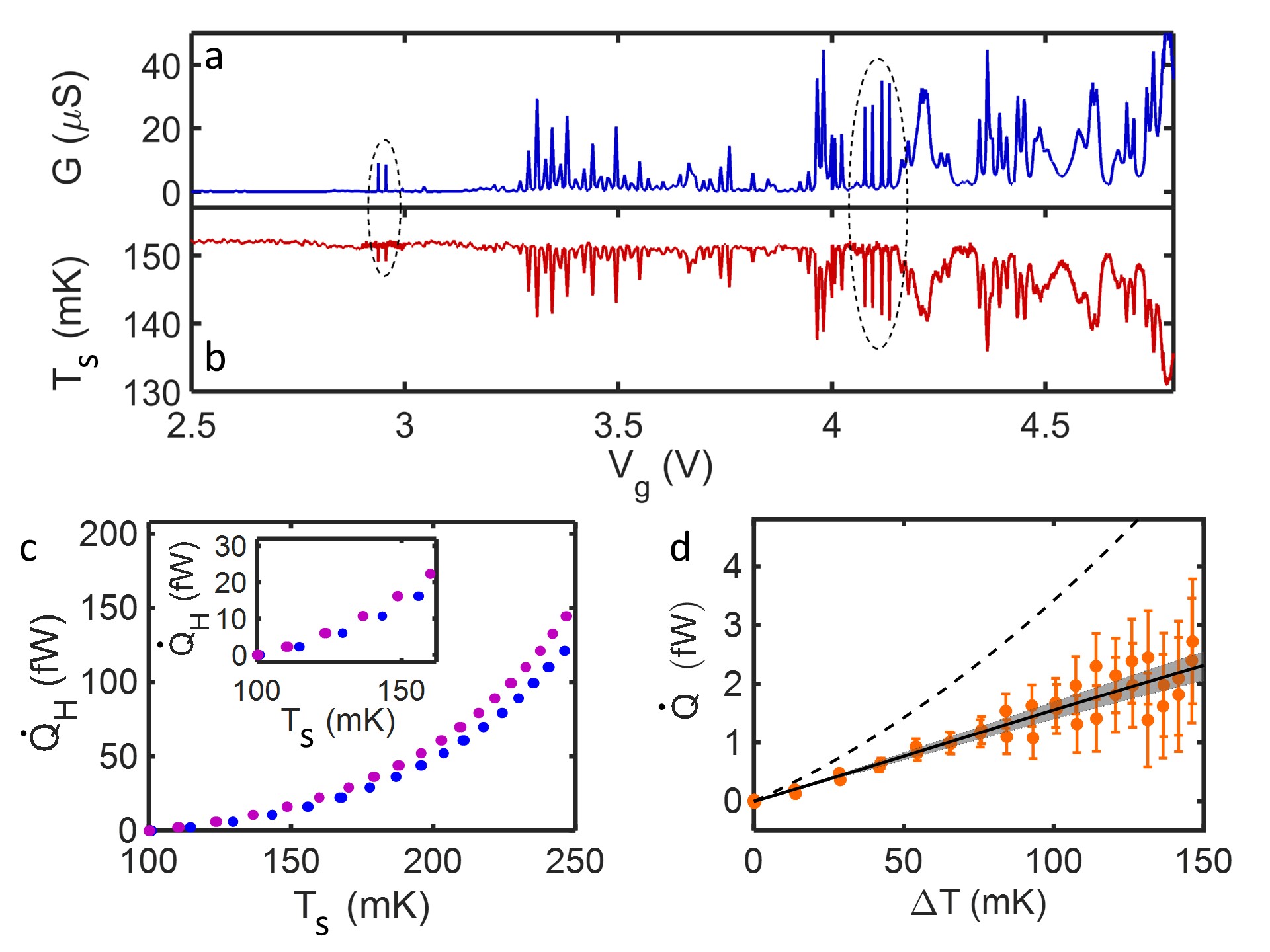}
	%\centering
 \vspace{-0.5cm}
	\caption{Gate dependence of (a) the linear charge conductance of the InAs nanowire at thermal equilibrium, and of (b) the temperature of the overheated source island. (c) Heat injected in the source on (red) and off (blue) a conduction resonance near $V_g=2.9$ V, characterized by $\Gamma/(k_BT)\approx 7$. (d) Heat current across the nanowire deduced from the difference of the two preceding measurements (orange). The dashed and the continuous lines show the expected behavior from the Wiedemann-Franz law and the scattering approach, respectively. All data are taken with $T_b=100$ mK. Adapted with permission from Ref. (\onlinecite{majidi2022quantum}). Copyright 2022 American Chemical Society.}
	\label{Nw+heatvalve}
\end{figure}

\subsubsection{Energy-selective transmission}
Early reports on quantum point contacts (QPCs) showed that the heat conductance of individual quantum conductance channels qualitatively followed the charge conduction pattern, as the QPC was progressively opened \cite{molenkamp1992, prance2009electronic}. This observation was followed by the quantitative demonstration that the opening of one supplementary quantum conduction channel provided an increase of heat conductance given by the WF law within less than a few percent. This is expected because $T(E)$ is flat in between two conduction thresholds in the QPC gate space. It also indicates that there are no interaction effects at play (see below for a refinement of this discussion). 

In the absence of interactions, the crucial role of $T(E)$ in deviations from the WF law, as discussed in Section II, was observed in single quantum dot junctions. A first study based on electromigrated quantum dot junctions demonstrated good qualitative agreement with the predictions of NEGF theory (Section II.A), however, the tunnel couplings were too large to be accurately determined independently \cite{dutta2020single}. Shortly after, experiments using a quantum dot formed in an InAs nanowire allowed smaller and better characterized tunnel couplings, with measurements that were in quantitative agreement with theory, based on the Landauer-B\"uttiker scattering approach, as described in Section II.C \cite{majidi2022quantum}. Clearly, deviations from the WF law appear most strongly when moving to the weak-coupling regime, in which $k_BT$ is larger than the spectral broadening of $T(E)$, given here by the $\hbar \Gamma$ broadening of the tunnel coupled quantum level. This is visible in \mbox{Fig.} \ref{Nw+heatvalve}d, where the condition $\Gamma\approx 7 k_BT$ is sufficient to induce a heat conductance about 35 \% below the WF law in the linear regime. Notably the deviation is even better seen in the non-linear regime of temperature gradients. For larger values of $\Gamma/(k_BT)$, the WF law is again recovered to a very good approximation. The suppression of the heat conductance {\em below} the WF prediction by a narrow conduction resonance can be understood intuitively as stemming from the pass-band energy filter provided by the quantum level: only particles inside an energy window $\sim \Gamma$ are allowed to tunnel back and forth. The energy exchanged per tunneling event is thus limited by $\Gamma$ and not $k_BT$. 

\subsubsection{Interactions}
\begin{figure}[t!]
	\includegraphics[width=\columnwidth]{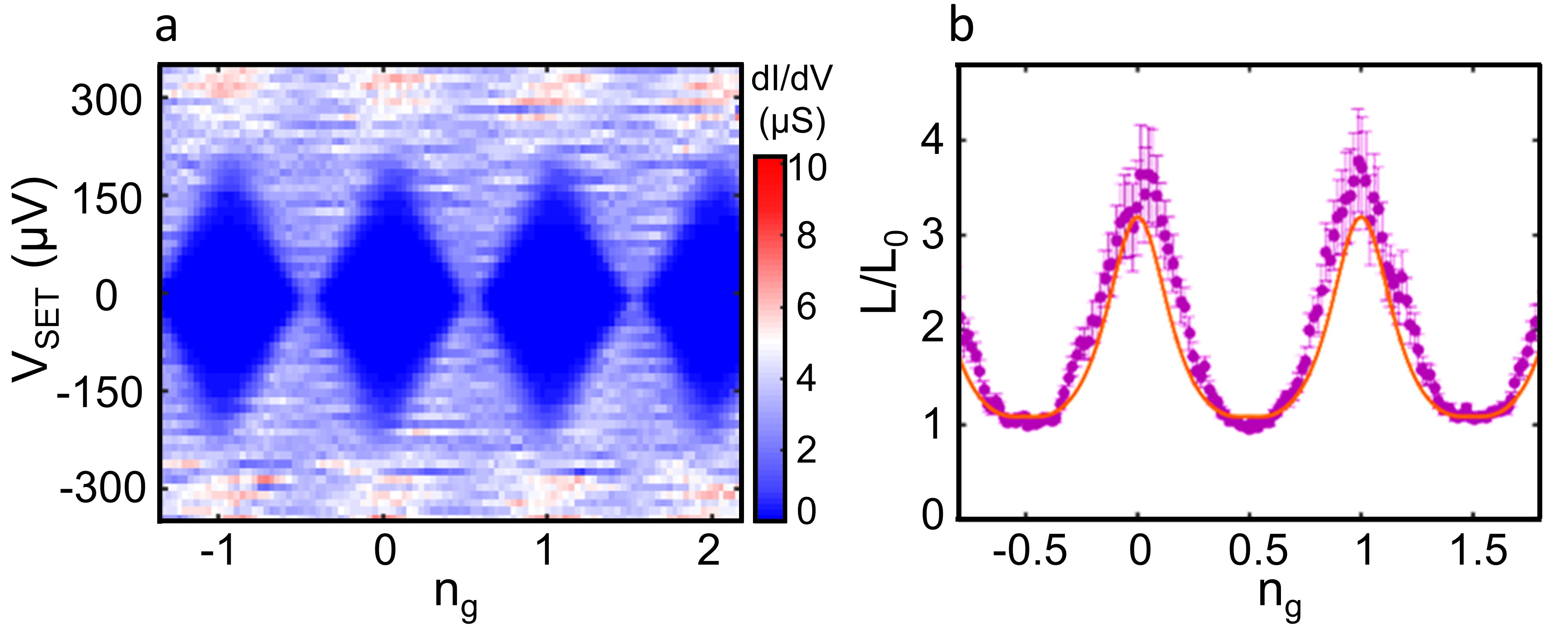}
	\centering
	\caption{ (a) Differential conductance map of a single-electron transistor, as a function of drain-source voltage and the gate-driven average electronic occupation number (with an arbitrary integer offset), displaying Coulomb diamonds. (b) Normalised heat conductance $L/L_0$ of the same device (purple dots) together with a theoretical calculation (line). Data taken at $T_b=160$ mK. Adapted with permission from Ref. (\onlinecite{dutta2017thermal}). Copyright 2017 American Physical Society.}
	 \label{bivas}
\end{figure}

As discussed in Section II, treating the combination of energy-selective tunneling processes and interactions is not an easy theoretical task. In the weak-coupling regime however, such that the energy scale $\hbar \Gamma$ can be treated perturbatively. The effect of interactions was calculated analytically in the SET geometry using a diagrammatic approach based on the QME formalism described in Section II.B \cite{kubala2008violation}. Experiments based on a SET with charging energy $\sim$150 $\mu$eV (\mbox{Fig.} \ref{bivas}a showed a strong enhancement by up to 300\% of the Lorentz ratio $L/L_0= \kappa/(L_0GT)$ when entering the Coulomb blockaded regime \cite{dutta2017thermal}, in excellent agreement with theory (\mbox{Fig.} \ref{bivas}b). An intuitive way of understanding the relative enhancement of heat conduction above the WF prediction is by considering the Coulomb blockaded island as a high-pass energy filter for transport processes. In contrast with the non-interacting resonant transmission discussed above, the tunneling of high-energy particles is here allowed and leads to an average energy transfer larger than $k_BT$ per tunneling event.

A particularly intriguing regime can be observed in the case where the source island - rather than the device itself - is subject to Coulomb blockade. Early studies based on quantum dots in a two-dimensional electron gas already noted the detrimental consequences of charging effects in one of the heat reservoirs, on the ability of the quantum dot device to conduct heat \cite{prance2009electronic}. It was later predicted \cite{slobodeniuk2013} and experimentally confirmed \cite{sivre2018heat} that for a Coulomb blockaded source reservoir, the reduced heat conductance of $N$ fully transmitting conduction channels was actually $L/L_0=(N-1)/N$. This can be interpreted as a single out of the $N$ conduction channels not contributing to heat transport, due to the suppression of energy fluctuations in the source island. In the heat conductance experiments using fully metallic bolometers \cite{dutta2017thermal,dutta2020single,majidi2022quantum}, care was taken to avoid possible Coulomb blockade effects in the source by using at least one transparent contact to the bolometer.

\subsection{Thermopower}

 We now move to the thermoelectric signatures of electronic transport across a quantum dot junction, that is, the appearance of a voltage drop at open circuit conditions in response to a temperature gradient across the junction, as defined in \mbox{Eq.} (\ref{eq:S}). The weak-coupling (sequential tunneling) limit across a single quantum level is immediately seen from the Onsager relations to be a linear function of the level position, which is odd with respect to the electron-hole symmetric point (Fermi level). This leads to a characteristic sawtooth-shaped thermoelectric signal $S(V_g)$ as a function of the gate voltage controlling the position of the relevant energy level \cite{beenakker1992theory,staring1993coulomb}. Cotunneling contributions essentially quench the thermoelectric response as soon as these become dominant over sequential tunneling, as occurs for instance deep inside a Coulomb diamond \cite{turek2002cotunneling}. However, cotunneling does not affect the odd electron-hole symmetry of $S$. Quantum dot junctions have been theorized \cite{mahan1996} and experimentally demonstrated \cite{humphrey2005,josefsson2018quantum} to allow for very high thermoelectric efficiencies, that is, work extraction from the heat current imposed by temperature gradient. 

 \begin{figure}[t!]
 \begin{centering}
 	%\hspace{-2.5cm}
	\includegraphics[width=\columnwidth]{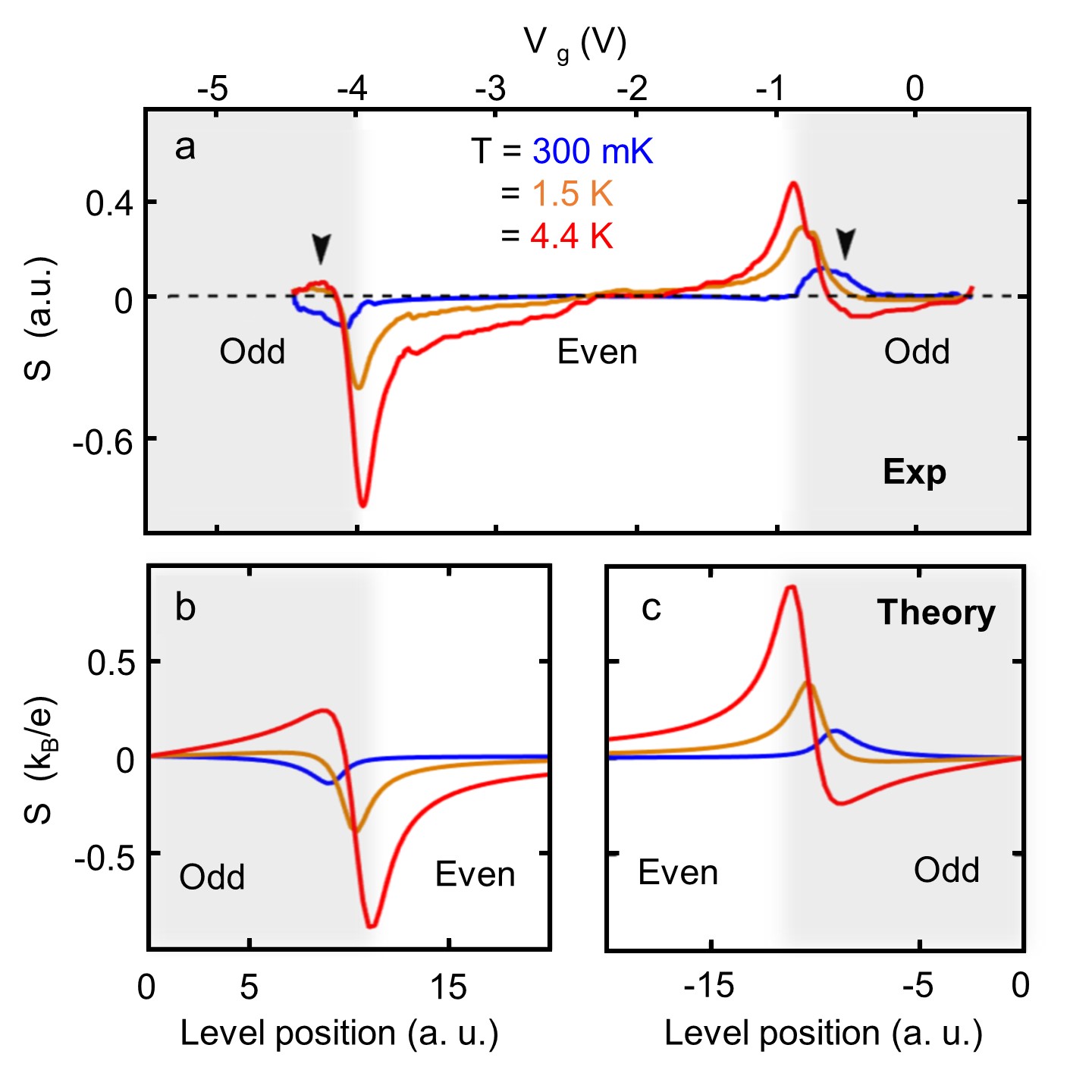}     
 \end{centering}
  	\centering
 \vspace{-0.1cm}
	\caption{(a) Experimental thermopower $S(V_g)$ at three experimental device temperatures of 300 mK (blue), 1.5 K (orange) and 4.4 K (red). The arrows highlight the level depths in the Kondo regime near which the thermopower changes sign with increasing temperature. (b,c) Corresponding numerical renormalization group calculation using the same set of temperatures $T$ (with the same color code) and microscopic parameteres extracted from the charge transport measurements. Adapted with permission from Ref. (\onlinecite{dutta2018direct}). Copyright 2019 American Chemical Society.}
	\label{thermopower}
\end{figure}

At higher tunnel couplings, quantum dot junctions can form complex many-body ground states with the leads, which manifest through the appearance of the so-called Abrikosov-Suhl (AS) resonance in the spectral function near the Fermi energy of the contact. This resonance is observable as a zero-bias conductance peak (Kondo effect) in quantum dot junctions  \cite{goldhaber1998kondo} under a degeneracy condition, which is for instance realised by the spin-1/2 doublet of a singly occupied level. Numerical renormalization group calculations predicted that in such a device $S(V_g)$ would have a markedly different behavior in adjacent Coulomb diamonds, corresponding to opposite occupation parities \cite{costi2010thermoelectric}. This leads thus to a $2e$-periodic signal in the charge state of the quantum dot and marked deviations from the above-discussed asymmetric thermopower signal as $V_g$ crosses a charge degeneracy point. Furthermore, the rapid collapse of the AS spectral resonance with increasing temperature must lead to a characteristic sign change of the Kondo-thermoelectric signal, as the spectral weight of the AS resonance decreases below the contribution from cotunneling \cite{costi2010thermoelectric}. These predictions were confirmed in detail by two recent experimental studies \cite{svilans2018thermoelectric,dutta2018direct}, which used InAs nanowires and Au nanoparticles as the quantum dot, respectively. \mbox{Fig.} \ref{thermopower} highlights for instance the markedly different thermoelectric response at two consecutive charge degeneracy points as well as its sign change with increasing temperature in the odd occupation sector are highlighted, in agreement with predictions. 

\section{Implications for quantum technologies}

In conclusion, our discussion reviewed how electron interactions along with quantum confinement in nanoscale conductors can lead to significant thermoelectric effects and marked deviations of the electronic heat conduction as compared to the WF estimation. For instance, the heat conductance of a weakly coupled single quantum level is lower by about $\Gamma/(k_BT)$ as compared to the WF result. For thermoelectric applications, this could be a great benefit as it could allow - at least in theory - extremely high efficiencies \cite{mahan1996, humphrey2005,josefsson2018quantum}. For heat draining applications, one must however bear in mind the possibly drastic reduction of heat conduction by the leads, in case these are small enough for inducing for instance lateral quantum confinement.

From a basic science perspective, several unanswered questions remain. For instance, the thermal conduction in the presence of Kondo correlations \cite{boese2001thermoelectric,costi2010thermoelectric} has not yet been experimentally addressed. Furthermore, while the mechanisms of heat conduction in the presence of phase coherence were already investigated by a series of experiments in superconducting circuits \cite{giazotto2012josephson,martinez2015rectification}, similar experiments in normal coherent conductors are still lacking.

On the application side, thermal effects play also a pivotal role in the performance of quantum systems across various platforms. Therefore, the understanding and control of heat capacities and thermal conductances in nanoscale quantum conductors is called to play a growing role with the increasing technological readiness of quantum devices. Because superconducting qubit circuits are rather extended objects, issues related to overheating, due for instance to microwave drives, have not yet been reported. Conversely, semiconducting spin-qubit devices are truly nanoscale, which makes them promising from the point of view of integration and scalability. However, owing to the impressive recent progress in the number of coupled physical qubits, it has become clear that overheating is now the main mechanism limiting the operation fidelity \cite{philips2022universal, undseth_hotter_2023, lawrie2023simultaneous}. Even small temperature fluctuations can modify the qubit resonance conditions and drastically enhance decoherence. 

Eventually, the accurate determination of the thermal properties of nanoscale conductors in the quantum regime will become crucial in the coming years \cite{pekola2021colloquium} in order to allow mitigating the effects of overheating in dense architectures of quantum devices. There is now a strongly growing interest in determining the  energetic cost of quantum computing \cite{auffeves2022}, both on the macroscopic level as well as from elementary considerations of quantum thermodynamics \cite{hofmann2017heat,elouard2017extracting, pekola2015towards}. It can be anticipated that the theoretical and experimental tools presented here will contribute to laying the foundations of this emerging field.

\section*{Acknowledgments}

We acknowledge insightful discussions with Boris Brun, Victor Champain, Denis Basko, Bivas Dutta, Nicola Lo Gullo, Jukka Pekola, and Rob Whitney. This work received support from the European Union under the Marie Sklodowska-Curie Grant Agreement No. 766 025, the Swedish Research Council (DNR 2019-04111), and Nano-Lund. JPB was graciously supported by the National Science Foundation under award number DMR-1809024.

The authors have no conflicts of interest to disclose.
The data of the figures presented in this work have been published elsewhere and can be obtained from the corresponding author upon reasonable request.

%\section*{References}
\bibliography{APL.bib}

%% graphical table of contents.
%%%%%%%%%%%%%%%%%%%%%%%%%%%%%%%%%%%%%%%%%%%%%%%%%%%%%%%%%%%%%%%%%%%%%
\end{document}